# Two-phonon coupling to the antiferromagnetic phase transition in multiferroic $BiFeO_3$

Mariola O. Ramirez[1], M. Krishnamurthi[1], S. Denev[1], A. Kumar[1], Seung-Yeul Yang,[2] Ying-Hao Chu[2], Eduardo Saiz[3], Jan Seidel[2], A.P Pyatakov[4], A.Bush[5], D. Viehland[6], J. Orenstein[7], R. Ramesh[2], Venkatraman Gopalan[1, a)]

[1]*Department of Materials Science and Engineering and Materials Research Institute, Pennsylvania State University, University Park, Pennsylvania 16802*

[2]*Department of Materials Science and Engineering and Department of Physics, University of California, Berkeley, California 94720-1760*

[3]*Materials Sciences Division, Lawrence Berkeley NationalLaboratory, Berkeley, California 94720*

[4]*Physics Department, Moscow State University, Leninskie gori, 38, Moscow 119992, Russia*

[5]*Moscow State Insitute of Radio Engineering, Electronics and Automation, Vernadskii Prospect, 78, Moscow 117454, Russia.*

[6]*Department of Materials Science and Engineering, Virginia Tech, Blacksburg, Virginia 24061, USA*

[7]*Department of Physics, University of California, Berkeley, California 94720-1760*

a)

A prominent band centered at ~1000-1300 $cm^{-1}$ and associated with resonant enhancement of two-phonon Raman scattering is reported in multiferroic $BiFeO_3$ thin films and single crystals. A strong anomaly in this band occurs at the antiferromagnetic Neel temperature, $T_N$ ~375 °C. This band is composed of three peaks, assigned to $2A_4$, $2E_8$, and $2E_9$ Raman modes. While all three peaks were found to be sensitive to the antiferromagnetic phase transition, the $2E_8$ mode, in particular, nearly disappears at $T_N$ on heating, indicating a strong spin-two phonon coupling in $BiFeO_3$.

Multiferroics, specially materials that combine spontaneous magnetic and ferroelectric order parameters, are currently the subject of intensive investigations because of their potential for electrical control of magnetism, and vice versa.[1-5] Bismuth ferrite, $BiFeO_3$ (BFO), is probably the most widely studied, since it is exhibits multiferroicity at room temperature, with a coexistence of ferroelectricity and antiferromagnetism up to its Neel temperature of $T_N$~ 375 °C. At room temperature (RT), bismuth ferrite is a rhombohedrally distorted ferroelectric perovskite with space group *R3c* and Curie temperature, $T_c$ ~ 830°C.[6] It also shows a *G*-type canted antiferromagnetic order below Neel temperature, $T_N$ ~ 375 °C. It is not homogeneous in space in single crystals, but rather exhibits an incommensurately space-modulated spin structure along the $(110)_h$.[7]

While ferroelectricity is relatively easier to study in this material, the probing of magnetism is more challenging.[8] In particular, the coupling between ferroelectricity and magnetism is of key significance. Several techniques such as diffraction experiments (x rays, electrons and neutrons),[9] second harmonic generation (SHG),[10,11] and recently Raman spectroscopy,[12,13] have been employed. Evidence of the strong magneto-electric coupling in this material has been demonstrated by the observation of electrical control of both ferroelectric and antiferromagnetic domains in BFO films at room temperature.[14] Temperature dependence studies of the SHG intensity has also shown an evident decrease when approaching Neel temperature, though separating the antiferromagnetic contribution to SHG still remains a challenge.[8] Last year, Haumont *et al.* reported pronounced phonon anomalies around $T_N$ due to phonons influenced by spin correlations.[12,13] The reported Raman spectra by Haumont *et al.* have not fully agreed with other theoretical and experimental Raman and IR studies on thin films, ceramics or bulk BFO single crystals, suggesting variability between different samples.[15-20] The selection rules for the Raman active modes in rhombohedral $R3c(C_{3v}^{6})$ symmetry predict only 13 active Raman phonons with $A_1$ and E symmetries, according to the irreducible representation, $\Gamma_{RAMAN/IR} = 4A_1 + 9E$. In polarized Raman scattering, the $A_1$ modes can be observed by parallel polarization, while the E modes can be observed by both parallel and crossed polarizations. Since all these modes fall in the frequency range below ~ 700 $cm^{-1}$, most of the Raman studies have focused in this region, with the subsequent lack of information at higher frequencies. In this letter, we present Raman spectra in the 100-2000 $cm^{-1}$ spectral range in thin films and bulk single crystals. In addition to the well-understood Raman features in the low frequency region, the spectra show a very prominent band at ~ 1000-1300 $cm^{-1}$, which we associate with two-phonon Raman scattering, strongly enhanced due to the resonance with the intrinsic absorption edge. The temperature dependence of the two-phonon contribution to the total Raman spectrum has been analyzed in the proximity of Neel temperature. Remarkable changes in both, intensity and spectral shape have been observed, pointing out the strong spin-two phonon coupling in BFO.

4.5 μm $BiFeO_3$ films on (110) $DyScO_3$ were grown by metalorganic chemical vapor deposition equipped with a liquid delivery system. Triphenylbismuth [Bi(Ph)3] and tris(2,2,6,6-tetramethyl-3,5-heptanedionate)iron [Fe(thd)3] dissolved in tetrahydrofuran were used as the liquid metalorganic precursor materials. The supply rates for the bismuth and iron sources were 7.8×10−6 and 1.22×10−6 mol/min, respectively, and the substrate was held 620 °C during deposition. The thick films were relaxed, and have $(001)_p$ pseudocube-on-psudocube epitaxial geometry, with trigonal $C_{3v}$ crystal structure. $BiFeO_3$ single crystals with $(001)_p$ crystal faces were grown in a sealed platinum crucible from a flux melt of $Bi_2O_3$, $Fe_2O_3$, and NaCl (75.6:17.9:6.5 molar ratio). The flux was cooled during several days from 900°C to 820°C at a rate of 0.5° per hour. The crystals were removed from the flux by dissolving mechanically broken flux parts in diluted nitric acid. Raman spectra were recorded in a back-scattering geometry by using a WITec alpha 300 S confocal Raman microscope. A grating with 600 lines/mm blazed at 500 nm was used to provide spectral information in a high frequency range. Due to the considerable absorption of BFO at 514 and 488 nm excitation wavelengths, low laser excitation power density (~ 1mW/area) was employed to avoid significant heating effects. Temperature measurements up to 550 ºC were carried out by using a commercial LINKAM heating stage placed under the Raman microscope; The heating rate was 5 ºC/min. For the $(001)_p$ crystallographic growth planes studied here, the optic axis is at an angle to the surface, and polarized Raman signal mixes multiple polarizability tensor components; hence Raman spectra were collected in unpolarized geometry.

Figure 1(a) and (b) shows the unpolarized Raman spectra as a function of temperature under excitation at 514 nm, for the $BiFeO_3$ thin film and single crystal, respectively. As seen, both the thin films and the single crystal show the same Raman modes at similar energy positions. When comparing the low frequency Raman modes in Figure 1 (100-700 $cm^{-1}$), with those previously theoretical and experimentally reported, good agreement is obtained. The only difference is the prominent additional band around ~1000-1300 $cm^{-1}$ which has not been reported before. Changing the excitation wavelength to 488nm led to no observed spectral shifts in this band; therefore, its Raman scattering nature is confirmed, excluding any possible phonon and/or magnon assisted luminescence in this region. The origin of this structure has been assigned to the combination of three different two-phonons Raman scattering in BFO labeled as $2A_4$, $2E_8$ and $2E_9$, since their spectral positions correspond to practically double the energy values of the $A(LO_4)$ ~480 $cm^{-1}$, $E(TO_8)$ ~550 $cm^{-1}$ and $E(TO_9)$ ~620 $cm^{-1}$ normal modes of BFO respectively.[15,16] The strong contribution of the two-phonon band to the total Raman spectrum has been attributed to a resonant enhancement with the intrinsic absorption edge in BFO (2.66 eV). This is similar to the 2-phonon bands reported in hematite, $\alpha$-$Fe_2O_3$, the simplest case of iron oxides containing only $FeO_6$ octahedra.[21,22] This resonant enhancement also explains the intensity increase observed when the excitation wavelength was tuned to 488 nm. Furthermore, if we use the striking spectral similarity between BFO and $\alpha$-$Fe_2O_3$, the two phonon scattering band observed in BFO can be directly correlated with the strong band found at 1320 $cm^{-1}$ in hematite, previously wrongly assigned as two magnon scattering, and later on identified as a two-phonons (~620 $cm^{-1}$) overtone.[21,23] As the samples are heated (Fig. 1), the Raman modes gradually broaden, as well as slightly shift to lower wavenumbers, which is expected due to thermal expansion and thermal disorder respectively. The most striking feature, however, is a dramatic decrease in the total integrated intensity, as well as the spectral shape of the two-phonon Raman band with increasing temperature.

Figure 2 shows a detail of the Raman spectrum in the 700-1800 $cm^{-1}$ wavenumber region below (RT) and above (400 °C) the antiferromagnetic phase transition in BFO. In both cases, the broad two-phonons overtone can be fitted into three Gaussian bands peaking at around 968, 1110 and 1265 $cm^{-1}$ at room temperature and 961,1092 and 1258 $cm^{-1}$ at 400 °C, i.e practically the double energy values of $A_4$, $E_8$ and $E_9$ normal modes in BFO. From the fit, the full width at half maximum (FWHM) of these bands was found to be 115, 123 and 145 $cm^{-1}$, respectively at RT and 145, 139 and 170 $cm^{-1}$ at 400 °C. The spectral shifts and broadening occur due to thermal dependence of Raman modes. At first sight, on comparing both spectra, a clear change in the spectral shape is observed, mainly due to the strongly reduction of the $2E_8$ contribution when passing through the antiferromagnetic phase transition.

Figure 3 shows the total contribution to the Raman spectra of the most intense $2E_9$ two-phonon band (center at ~1260 $cm^{-1}$) as a function of sample temperature. In this figure the intensity of the two-phonons scattering $I_{2P}$ has been normalized to the one- phonon scattering intensity, $I_P$, of the Raman active mode at 254 $cm^{-1}$ and reduced by the appropriate thermal population factor

$$R = I_{2P}(n(\omega_p)+1) \big/ I_p(n(\omega_{2p})+1)^2 \quad (1)$$

where $n(\omega)=[e^{h\omega/kT}-1]^{-1}$ is the Bose-Einstein factor. By using the reduced Raman intensity given by *R,* the contribution of the Bose-Einstein population from the measured Raman intensity is eliminated, and the intensity changes can be compared independent of the population considerations. On approaching the Neel temperature, $T_N$, a remarkable decrease in the temperature dependence of the integrated intensity of the two-phonon Raman scattering in $BiFeO_3$ is seen, followed by a constant value after the antiferromagnetic phase transition. Similar results were obtained when the integrated intensity corresponding to the other two phonons overtone center at 965 and 1110 $cm^{-1}$ was analyzed. Therefore, a strong interplay between the ferroelectric and magnetic subsystems of BFO can be concluded. We particularly note the high sensitivity of the two-phonon band to the antiferromagnetic phase transition in BFO as compared to the one-phonon scattering mode, since its total contribution to the Raman spectrum was referred (normalized) to the one-phonon band. Figure 3 also shows the ratio between $2E_8$ and $2E_9$ two-phonons scattering integrated intensity $\Gamma(2E_8)/\ \Gamma(2E_9)$, as a function of temperature. As can be seen, it remains almost constant up to ~200 ºC and abruptly decreases in the vicinity of $T_N$, supporting again the significant spin-two phonon coupling in BFO.

A plausible model to explain this behaviour may relate to the specific bond motion of the Raman modes and the octahedral rotation along the [111] axis responsible of the weak ferromagnetism in BFO.[24] Previous studies on the assignment of the Raman modes in BFO thin films with pseudo-tetragonal symmetry attributes the $A_1$ modes and the low frequency E modes modes (< 400 $cm^{-1}$) to Bi-O1 bonds. The higher frequency E modes are attributed to Fe-O bonds. Most specifically, some of them are related to Fe-O1 and others with Fe-O2, where O1 are axial and O2 are equatorial ions.[17] This assignment also agrees with other Raman scattering studies on $Bi_{1-x}Nd_xFeO_3$ multiferroic ceramics where a change of Bi-O covalent bonds is observed with increasing $x$.[15] Therefore, from the results displayed in Figure 2, we can tentatively assign the $2E_8$ overtone to Fe-O1 bonding and the $2E_9$ to Fe-O2 and hence relate it with the octahedral rotation critical to weak magnetism. Oxygen rotation plays an important role in the antiferromagnetism through superexchange, which is very dependent on bond angles. Thus, if there is spin-phonon coupling, the structural distortions due to the G-type canted antiferromagnetic order should be reflected in the evolution of Raman scattering spectra, which might explain the different evolutions of the two-phonons overtones on approaching the Neel temperature (See Fig. 3). We also note a recent work that reassigns the Raman mode at 550 $cm^{-1}$ as $A_1$(TO) transversal mode since it was only observed in parallel polarized pump and Raman signals geometry.[25] Though we cannot ascertain the new assignment in this study, the plausible explanation proposed could also account for this new assignment since a slight spin-phonon coupling to the antiferromagnetic phase transition is predicted for both the $A_1$(TO) and E(TO) modes by first principle calculations.[18,26]

In summary, a resonant enhancement of two-phonons Raman scattering in the vicinity of 1200 $cm^{-1}$ has been reported in BFO multiferroic system. Temperature studies well above the Neel temperature shows a strong coupling of the two-phonon band to the antiferromagnetic phase transition in BFO. Significant changes in the integrated intensity as well as in the spectral shape when approaching $T_N$ were observed. Additionally, since no dramatic changes - excepting for the intensity variations above mentioned- were observed in the overall Raman spectrum when crossing the Neel temperature, it is possible to conclude that the antiferromagnetic phase transition in BFO does not seem to be accompanied by dramatic structural changes. From the results presented in this work, it is clear that spin-phonon coupling, as well as the temperature dependence of the ferroelectric order parameter cannot be neglected in the proximity of $T_N$.

We acknowledge funding from the National Science Foundation grant numbers DMR-0512165, DMR-0507146, DMR-0213623, and DMR-0602986.

## REFERENCES

**1.-** T. Kimura, T. Goto, H. Shintani, K. Ishizaka, T. Arima and Y. Tokura Nature, (London) **426**, 55 (2003).

**2.-** N. Hur, S. Park, P.A. Sharma, J.S. Ahn, S. Guha and S.W. Cheong, Nature, (London) **429**, 392 (2004).

**3.-** N.A Spaldin and M. Fiebig, Science, **309**, 391 (2005).

**4.-** W. Erenstein, N.D. Mathur, and J.F. Scott, Nature (London), **442**, 759, (2006).

**5.-** M. Fiebig, Th. Lottermoser, D. Frohlich, A. V. Goltsev and R.V. Pisarev. Nature (London), **419**, 818, (2002).

**6.-** F. Kubel and H. Schmid, Acta Cryst. B, **46,** 698, (1990).

**7.-** P. Fischer, M. Polomska, I. Sosnowska and M. Szymanski, J. Phys. C. **13**, 1931, (1980).

**8.-** M. Fiebig, V.V. Palov and R.V. Pissarev, J. Opt. Soc. Am. B, **22**, 96, (2005).

**9.-** A. Palewicz, R. Przenioslo, I. Sosnowska and A.W. Hewat, Acta Cryst. B, **63**, 537, (2007).

**10.-** A.M. Agaltsov, V.S. Gorelik, A.K. Zvezdin, V.A. Murashov and D.N. Rakov, Sov. Phys. Short Commun. **5**, 37, (1989).

**11.-** M.S. Kartavtseva, O.Yu. Gorbenko, A.R. Kaul, T.V. Murzina, S.A. Savinov and O.A. Aktsipetrov, J. Mater. Res. **22**, 2063, (2007).

**12.-** R. Haumont, J. Kreisel, P. Bouvier and F. Hippert, Phys. Rev. B **73**, 132101, (2006).

**13.-** R. Haumont, J. Kreisel and P. Bouvier, Phase Transitions, **79**, 1043, (2006).

**14.-** T. Zhao, A. Scholl, F. Zavaliche, K. Lee, M. Barry, A. Doran, M.P. Cruz, Y.H. Chu, C. Ederer, N.A. Spaldin, R.R. Das, D.M. Kim, S.H. Baek, C.B. Eom and R. Ramesh, Nat. Mat, **5**, 823, (2006).

**15.-** G.L. Yuan, S.W. Or and H.L. Wa Chan, J. Appl. Phys. **101**, 064101, (2007).

**16.-** M.K. Singh, H.M. Jang, S. Ryu and M. Jo, Appl. Phys Lett. **88**, 042907, (2006).

**17.-** M.K. Singh, S. Ryu and H.M. Jang, Phys. Rev. B **72**, 132101, (2005).

**18.-** P. Hermet, M. Goffinet, J. Kreisel and Ph. Ghosez, Phys. Rev. B **75**, 220102, (2007).

**19.-** H. Fukumura, H. Harima, K. Kisoda, M. Tamada, Y. Noguchi and M. Miyayama, J. Magn. Magn. Mater. **310,** e367, (2007).

**20.-** S. Kamba, D. Nuzhnyy, M. Savinov, J. Sebek, J. Petzelt, J. Prokleska, R. Haumont and J. Kreisel, Phys. Rev. B **75**, 024403, (2007).

**21.-** K.F McCarty, Solid State Commun. **68**, 799, (1988).

**22.-** T.P. Martin, R. Merlin, D.R. Huffman and M. Cardona, Solid State Commun. **22**, 565, (1977).

**23.-** M.J Massey, U. Baier, R. Merlin and W.H. Weber, Phys. Rev. B **41**, 7822, (1990).

**24.-** C. Ederer and N.A Spaldin, Phys. Rev. B **71**, 060401, (2005).

**25.-** M. Cazayous, D.Malka, D. Lebeugle and D. Colson, Appl. Phys. Lett. **91**, 071910, (2007).

**26.-** C. Fennie and C. Ederer, *Private communication.*

## FIGURE CAPTIONS

**Figure 1.- (Color on line)**Temperature-dependent (Room temperature to 550 °C) unpolarized Raman scattering spectra of $BiFeO_3$ film **(a)** and bulk single crystal **(b)** recorded under excitation at 514 nm.

**Figure 2.- (Color on line).** Detail of the Raman spectrum in the 700-1800 $cm^{-1}$ wavenumber region. Solid lines are fits to three Gaussian functions corresponding to two-phonons replica of the $A_4$ ~483 $cm^{-1}$, $E_8$ ~550 $cm^{-1}$ and $E_9$ ~620 $cm^{-1}$ normal modes in BFO respectively. (These modes have been marked in Figure 1 (a)). **(a)** Room temperature **(b)** T = 400 °C.

**Figure 3.- (Color on line). (Left axis)** Temperature dependence of the reduced integrated Raman intensity of the 1260 $cm^{-1}$ two-phonons Raman scattering band in $BiFeO_3$ normalized to scattering intensity of the 254 $cm^{-1}$ phonon. A correction has been made to remove the dependence on thermal population factors. **(Right axis)** Integrated intensity ratio between $2E_8$ and $2E_9$ two-phonons Raman scattering, $\Gamma(2E_8)/\ \Gamma(2E_9)$, as a function of the sample temperature.

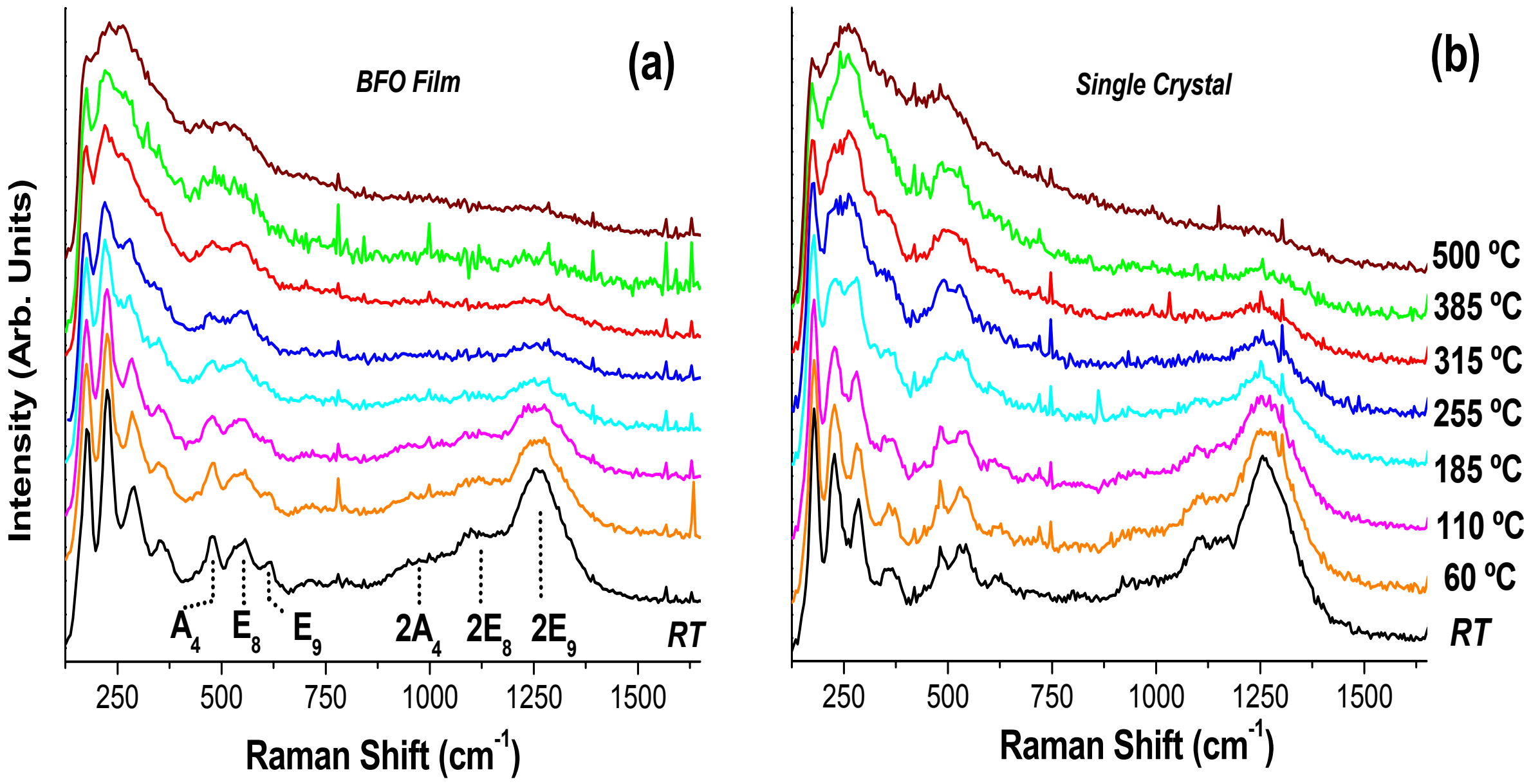

# FIGURE 1

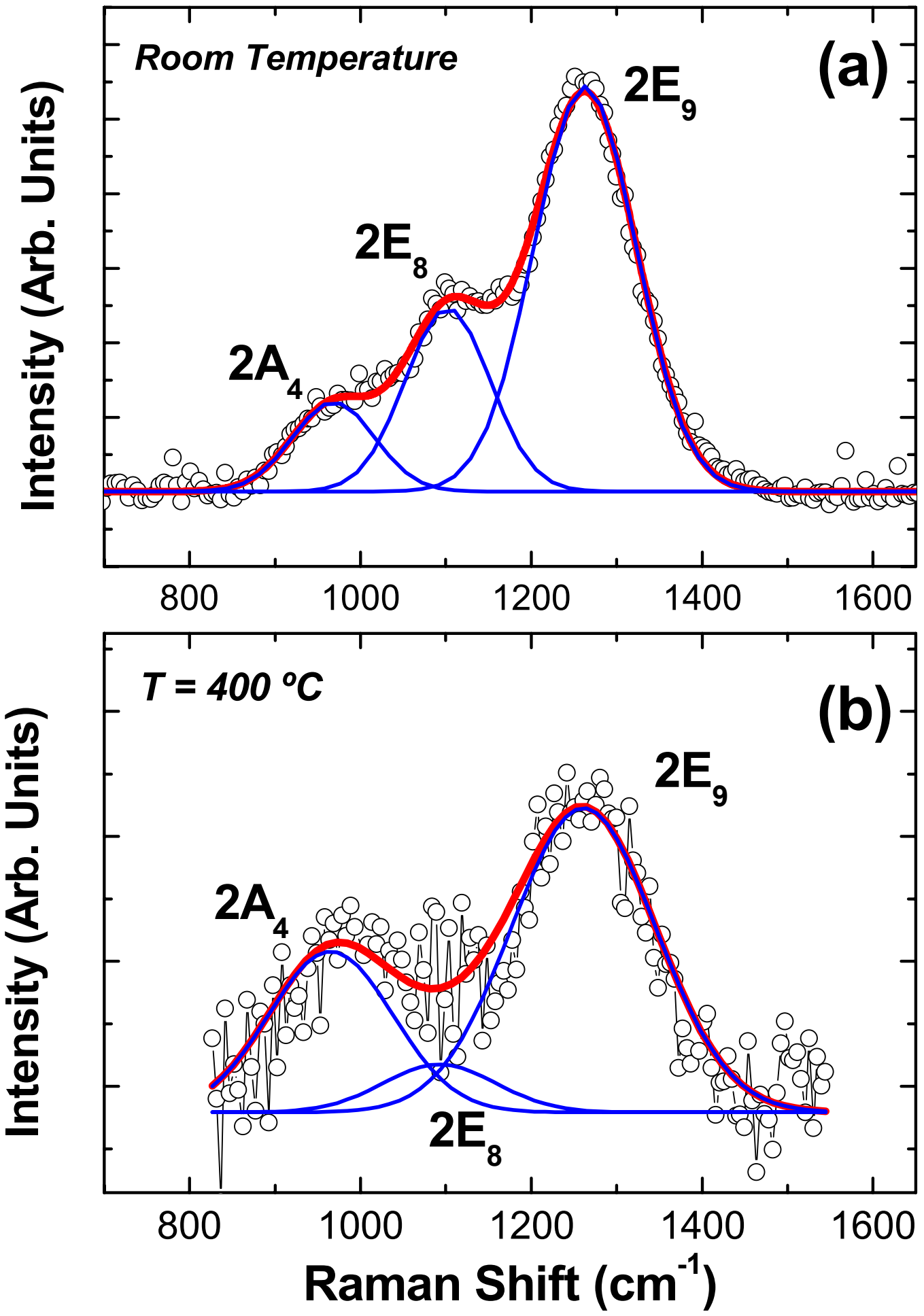

# FIGURE 2

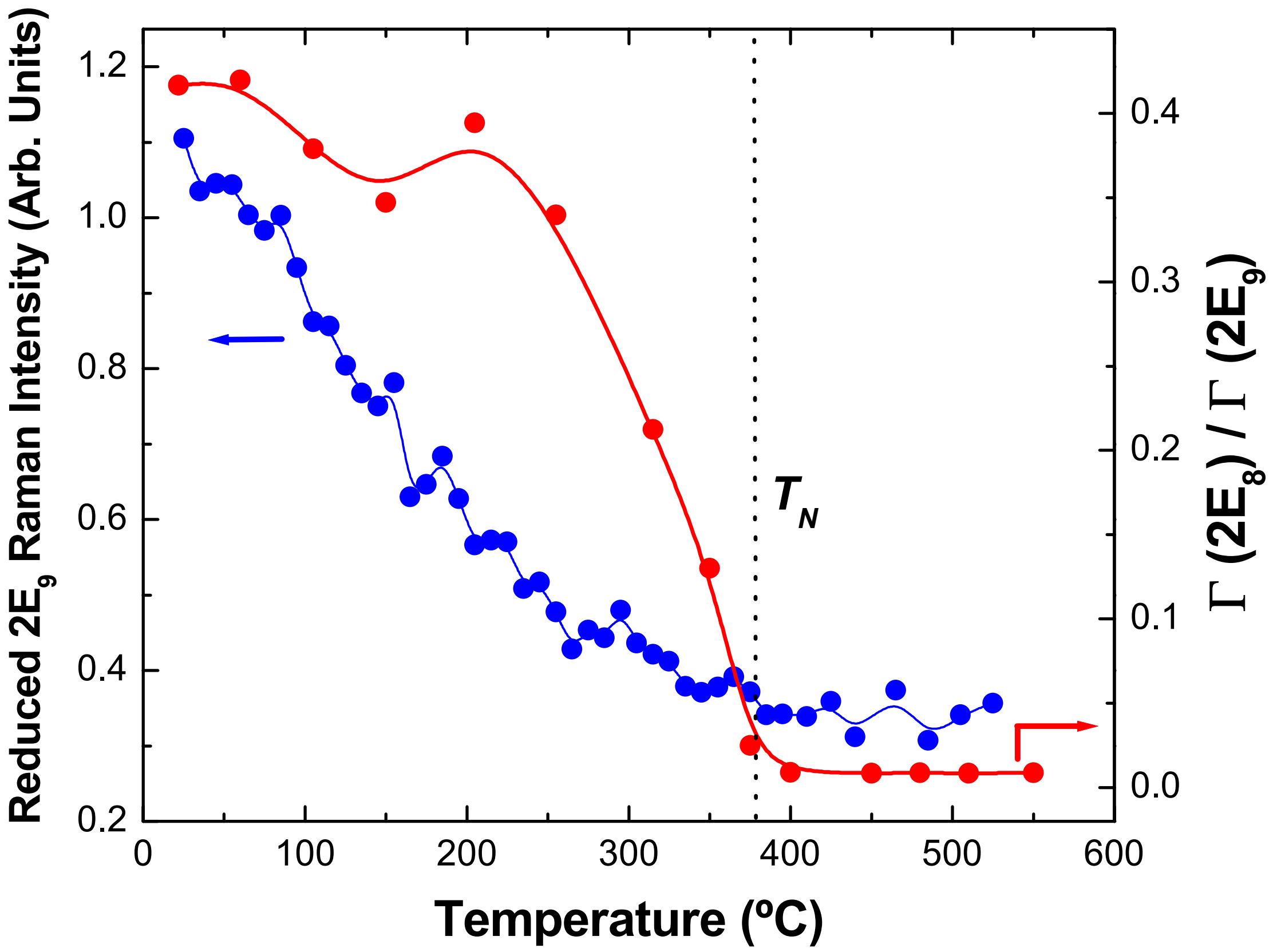

**FIGURE 3**